\documentstyle{l-aa}

\def\Msol{\,M_\odot}

\def\kms{kms$^{-1}$}
\def\vt{$v_{\rm t}$}

\def\teff{$T_{\rm eff}$~}
\def\C12C13{$^{12}$C/$^{13}$C}

\def\sixt`enth{\textstyle{1\over{16}}}
\begin{document}
\input{psfig.tex}

% !! Testo Latex

\thesaurus{ 7(08.06.3 - 08.01.3 - 08.01.1 - 08.16.3) }
\title{ Abundances of light elements in metal-poor stars. I. }
\subtitle{ Atmospheric parameters and a new \teff\ scale }
\author{ R.G. Gratton$^1$, E. Carretta$^{2,3}$, F. Castelli$^4$ }
\offprints{ R.G. Gratton }
\institute{ $^1$Osservatorio Astronomico di Padova, Vicolo dell'Osservatorio, 5,
	       I-35122 Padova, ITALY\\
	    $^2$Dipartimento di Astronomia, Universit\`a di Padova, Vicolo 
	       dell'Osservatorio 5, I-35122 Padova, ITALY\\
	    $^3$Osservatorio Astronomico di Bologna, Via Zamboni 33, 
	       I-40126 Bologna, ITALY\\
	    $^4$Osservatorio Astronomico di Trieste, Via G.B. Tiepolo, 11,
	       I-34131 Trieste, ITALY }
\date{}

\maketitle

\begin{abstract}
We present atmospheric parameters for about 300 stars of different chemical
composition, whose spectra will be used to study the galactic enrichment of Fe
and light elements. These parameters were derived using an homogenous iterative
procedure, which considers new calibrations of colour-\teff\ relations for F, G
and K-type stars based on Infrared Flux Method (IRFM) and interferometric
diameters for population~I stars, and the Kurucz (1992) model atmospheres. We
found that these calibrations yield a self-consistent set of atmospheric
parameters for \teff$>4400$~K, representing a clear improvement over results
obtained with older model atmospheres. Using this \teff-scale and Fe
equilibrium of ionization, we obtained very low gravities (implying
luminosities incompatible with that expected for RGB stars) for metal-poor
stars cooler than 4400~K; this might be due either to a moderate Fe
overionization (expected from statistical equilibrium calculations) or to
inadequacy of Kurucz models to describe the atmospheres of very cool giants.
Our \teff\ scale is compared with other scales recently used for metal-poor
stars; it agrees well with those obtained using Kurucz (1992) models, but it
gives much larger \teff's than those obtained using OSMARCS models (Edvardsson
et al. 1993). This difference is attributed to the different treatment of
convection in the two sets of models. For the Sun, the Kurucz (1992) model
appears to be preferable to the OSMARCS ones because it better predicts the
solar limb darkening; furthermore, we find that our photometric \teff's for
metal-poor stars agree well with both direct estimates based on the IRFM, and
with \teff's derived from H$\alpha$\ wings when using Kurucz models. 

\keywords{  Stars: fundamental parameters - Stars: atmospheres -
Stars: abundances - Stars: population II }
\end{abstract}

\section {Introduction}

The determination of elemental abundances in metal-poor stars is a basic
constraint for models of the chemical evolution of our Galaxy, and provide
wealth of data about the history of star formation (see e.g. Wheeler et al.
1989). A very important r\^ole is played by C, N, O, Na, Mg, and Fe, which are
amongst the most abundant elements, likely produced in a variety of
astronomical sites. The determination of accurate abundances for these elements
in a large sample of stars of different metallicities, and their discussion
within the framework of galactic evolution is the main purpose of the present
series of papers. In the course of this investigation, we found it necessary to
discuss a number of important, related issues in order to obtain more reliable
results: some of them (e.g. the solar abundances and the applicability of the
adopted model atmospheres in abundance analyses) were treated in a parallel
study of the spectra of RR Lyrae stars at minimum light (Clementini et al.
1995), and in more depth by Castelli and Gratton (1996). The present paper is
devoted to the presentation of the adopted atmospheric parameters; these were
obtained using an iterative procedure which exploits both photometric and
spectroscopic data. The most relevant feature is a new, hopefully improved
calibration of colours against effective temperatures (\teff's). In other
papers of this series (and with the contribution of other authors) we will
present a discussion of non-LTE effects, using specially devoted statistical
equilibrium computations and a new empirical calibration of the collisional
cross-sections, and a discussion of the adopted abundance indices and of the
derived abundances within the framework of models of the galactic chemical
evolution. 

The colour-\teff\ calibrations are by themselves a very interesting output of
the present investigation: they have an important impact on a broad class of
topics, including e.g. abundance analyses and comparisons between theoretical
isochrones and cluster colour-magnitude (c-m) diagrams. Our new \teff\ scale is
close to that determined by King (1993: hereinafter K93); it is based on
empirical determinations for population~I stars, applying corrections for
non-solar metallicities drawn from the same Kurucz (1992) model atmospheres
used in the analysis. A quite extensive discussion of this \teff\ scale is
given in Sect.~5; we conclude that it gives reliable and consistent results for
\teff$>4600$~K, while either the \teff\ scale or the same Kurucz models seem
inadequate for stars cooler than 4400~K. 

\section {Program stars}

Before discussing the derivation of the atmospheric parameters adopted in our
analysis, a short presentation of the observational data used in this series of
papers is required. The original material consists in about 400 high
resolution ($R\sim 50,000$), high $S/N$\ ($>150$) spectra of 19 metal-poor
stars acquired with the Short Camera of the Coud\'e Echelle Spectrograph (CES)
at the ESO Coud\'e Auxiliary Telescope, La Silla, and with the Coud\'e
Spectrograph at the 2.7~m telescope of the McDonald Observatory. The stellar
sample is the same used by Gratton \& Sneden (1991, 1994: hereinafter GS1 and 
GS2) in order to derive abundances of Fe-group and $n-$rich elements.
However, the initial observations were carried out during ESO test time, kindly
made available by Dr S. D'Odorico: the available observing time forced us to
concentrate on the brightest ($V\leq 8$) southern metal-poor stars accessible
during the observing run. This sample included only 19 stars; furthermore, all
extremely metal-poor stars are giants, with the only exception of the subgiant
HD~140283. This sample is too small for the present purposes and important
biases are present. For this reason, we decided to increase it by adding data
(equivalent widths $EW$) from a number of literature sources, reanalyzing them
in the most homogeneous way possible: these additional data allow to better
understand the selection biases present in our data, and to obtain a sample
large enough for a statistically significant analysis of the results. 

The additional sources of $EW$s were selected from papers based on high
resolution, high $S/N$\ observational material; we only considered studies
giving $EW$s for both Fe~I and Fe~II lines, since we used the Fe
equilibrium of ionization to derive gravities: for this reason, only papers
dealing with data having a rather large wavelength coverage were considered.
The following works were then considered: 
\begin{enumerate}
\item Tomkin et al. (1992: hereinafter TLLS): analysis of high excitation 
 C and O permitted lines in 34 unevolved metal-poor stars;
\item Sneden et al. (1991) and Kraft et al. (1992) (hereinafter collectively
 SKPL): abundances of O from forbidden lines and Na from the doublet at
 6154-60~\AA\ in 27 field halo giants. 
\item Edvardsson et al. (1993: hereinafter E93): analysis of high excitation
 O, Na, and Mg lines in about 180 field dwarfs with metallicity [Fe/H]$>-1$.
 Their data were integrated with $EW$s for high excitation C lines from
 Clegg et al. (1981) and Tomkin et al (1995), and forbidden O lines from 
 Nissen \& Edvardsson (1992) in a smaller number of stars.
\item Zhao \& Magain (1990: hereinafter ZM90): abundances of Na and Mg in
 20 metal-poor dwarfs.
\end{enumerate}

Since there is some overlap amongst these different samples, on the whole, data
for almost 300 stars over a wide range in luminosity and metal-abundance are
considered. In the remainining part of this paper, we will discuss the
derivation of the atmospheric parameters used in the analysis of all this
material. 

\section {Derivation of atmospheric parameters}

\begin{table*}
\caption[ ]{Polynomial coefficients of the empirical colour-\teff\ calibrations
for population I stars (valid for $n<$colour$<m$). Two relations are given for
giants; the first one should be preferred for stars with colours smaller than
the following limits: $B-V$=1.48, $V-R$=1.10, $R-I$=0.94, $J-K$=0.92,
$V-K$=3.10, $b-y$=0.739. The second one should be preferred for stars with
colours larger than these limits. Errors are standard deviations from fits} 
\label{tab:1}
\vskip 12pt
\begin{tabular}{lccrrrrcc}
\hline\hline
    & Class & Stars &a$_0$ &a$_1$ & a$_2$ &a$_3$  & $n$    &  $m$ \\
\hline
\\
$B-V$ & III & 81 &    8843 & $-$6982.5 &   3961.8 &  $-$980.78 & 0.08 & 1.54\\
      &     &    & $\pm$95 &     401.7 &    530.5 &     206.09 &      &     \\
$B-V$ & III & 22 &  $-$228 &    7992.9 &$-$3466.5 &            & 1.20 & 1.64\\
      &     &    & $\pm$93 &    4281.2 &   1511.8 &            &      &     \\
$B-V$ &  V  & 33 &    8905 & $-$6730.4 &   3173.8 &  $-$552.07 & 0.12 & 1.35\\
      &     &    & $\pm$91 &     522.5 &    903.9 &     434.6  &      &     \\
\hline
$V-R$ & III & 81 &    9245 & $-$8392.3 &   3247.6 &   $-$85.49 & 0.10 & 1.23\\
      &     &    &$\pm$140 &     887.6 &   1473.8 &     722.71 &      &     \\
$V-R$ & III & 22 &    7060 & $-$3879.5 &   1046.7 &            & 0.89 & 2.03\\
      &     &    & $\pm$77 &     476.7 &    168.2 &            &      &     \\
$V-R$ &  V  & 29 &    9065 & $-$8067.3 &   3568.5 &  $-$131.30 & 0.07 & 1.17\\
      &     &    &$\pm$182 &    1328.0 &   2605.5 &    1431.50 &      &     \\
\hline
$R-I$ & III & 81 &    8824 &$-$11747.6 &   7920.2 & $-$1123.14 & 0.03 & 0.94\\
      &     &    & $\pm$75 &     373.9 &    649.0 &     400.52 &      &     \\
$R-I$ & III & 22 &    5932 & $-$3109.0 &    969.0 &            & 0.63 & 1.91\\
      &     &    & $\pm$88 &     389.3 &    159.3 &            &      &     \\
$R-I$ &  V  & 33 &    8764 &$-$11850.9 &   9310.1 & $-$2265.45 & 0.05 & 0.85\\
      &     &    & $\pm$90 &     545.3 &   1201.8 &     963.69 &      &     \\
\hline
$J-K$ & III & 68 &    8553 &$-$10456.7 &   9613.5 & $-$3886.47 & 0.17 & 0.97\\
      &     &    & $\pm$84 &    1202.7 &   2233.7 &    1260.56 &      &     \\
$J-K$ & III & 16 &   10071 &$-$10088.2 &   3843.5 &            & 0.85 & 1.22\\
      &     &    & $\pm$50 &    2073.6 &   1017.7 &            &      &     \\
$J-K$ &  V  & 19 &    9850 &$-$17774.9 &  23374.3 &$-$12170.60 & 0.11 & 0.83\\
      &     &    &$\pm$164 &    2887.6 &   6894.7 &    4901.1  &      &     \\
\hline
$V-K$ & III & 78 &    8992 & $-$2887.9 &    556.9 &   $-$40.02 & 0.19 & 3.67\\
      &     &    & $\pm$43 &      78.9 &     44.3 &       7.20 &      &     \\
$V-K$ & III & 13 &    6733 & $-$1136.0 &     99.3 &            & 3.10 & 6.01\\
      &     &    & $\pm$63 &     325.1 &     35.1 &            &      &     \\
$V-K$ &  V  & 33 &    8890 & $-$2782.4 &    523.1 &   $-$31.79 & 0.29 & 3.32\\
      &     &    & $\pm$60 &     136.2 &     97.1 &      19.10 &      &     \\
\hline
$b-y$ & III & 32 &    8736 & $-$9595.6 &   6135.2 & $-$1428.90 & 0.06 & 0.95\\
      &     &    &$\pm$103 &     781.4 &   1738.9 &    1108.00 &      &     \\
$b-y$ & III & 14 &   11754 &$-$15965.3 &   8172.8 &            & 0.70 & 0.96\\
      &     &    & $\pm$60 &    5430.3 &   3261.7 &            &      &     \\
$b-y$ &  V  & 30 &    8592 & $-$7044.5 &$-$2200.5 &    5248.20 & 0.06 & 0.79\\
      &     &    & $\pm$94 &     966.8 &   2786.9 &    2251.70 &      &     \\
\hline
\end{tabular}
%\picplace{4cm}
\end{table*}

\begin{figure*}
\psfig{figure=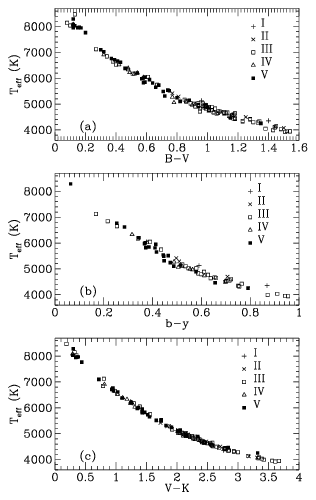,width=10cm,clip=}
\caption[ ]{Empirical relations between \teff and colours for the 140
population I stars with accurate \teff's from the IRFM. Stars of different
luminosity classes are marked with different symbols. For brevity, only
relations for $B-V$\ (panel a), $b-y$\ (panel (b), and $V-K$\ (panel c) are
shown in this figure} 
\label{fig:1}
\end{figure*}

When preparing the present series of papers, the grid of model atmospheres by
Kurucz (1992, hereinafter K92) became available to us\footnote{ 1993 CD-ROM 13
version of Atlas 9}. Since these models are able to reproduce various solar
features (flux distribution, photospheric abundances, limb darkening, etc.)
much better than the Bell et al. (1976; hereinafter BEGN) atmospheres, they
were adopted in the present analysis. We decided to revise the whole derivation
of atmospheric parameters (effective temperature \teff, surface gravity $\log
g$, model metal abundance [A/H], and microturbulent velocity \vt) for the
program stars, in order to put them on a self-consistent scale. \teff's were
derived from dereddened colours using semiempirical calibrations obtained
through a procedure similar to that of K93; however standard \teff's obtained
by means of the IRFM (Blackwell \& Shallis 1977) were taken from Blackwell \&
Lynas-Gray (1994: hereinafter BLG), who used new calibrations based on the K92
model atmospheres. Furthermore, corrections for gravities were also obtained
empirically by interpolating among values obtained for dwarfs and giants. The
adopted iterative procedure (which requires approximate initial values for
$\log g$, [A/H], and \vt) is as follows: 
\begin{enumerate}
\item We first obtained empirical colour-\teff's calibrations using a
compilation of \teff's derived with the IRFM for $\sim$140 population I stars
from the lists of BLG and Bell \& Gustafsson (1989); these last \teff's were
corrected to put them into the same scale of those of BLG. A first correction
is required because Bell \& Gustafsson derived \teff's from the IRFM using the
BEGN models rather than the K92 models used here; we notice that on this
respect, our calibrations represent an improvement over that of K93, who used
\teff's derived from the IRFM calibrated against BEGN models (Saxner \&
Hammarb\"ack 1985). The \teff's from Bell \& Gustafsson were then lowered by
122~K to put them on the same scale of those by BLG. This last correction might
be due to the different ways fluxes in the IR are determined by interpolation
within broad-band colours. Figure~\ref{fig:1} shows the colour-\teff(IRFM)
graphs. All colours are in the Johnson system, except the Str\"omgren colour
index $b-y$. The most appropriate colour indices can be deduced from the
dispersion of the observational data; as expected, the best index is $V-K$,
followed by $b-y$. However, it should be noticed that sequences for dwarfs
(luminosity class V; few stars of luminosity class IV are present in the
sample) and giants (luminosity class III) are clearly separated. For any
colour, we then fitted cubic polynomials
(\teff=$a_0+a_1~colour+a_2~colour^2+a_3~colour^3$) through the colour-\teff
planes for stars of luminosity class III and V separately. Polynomial
coefficients are given in Table~\ref{tab:1}. 
\item We derived \teff's from each observed colour for the program stars
using a cubic polynomial interpolation in the theoretical \teff-colour planes.
However, before doing this interpolation we transformed both theoretical
\teff's and observed colours, to consider the difference between the empirical
and theoretical calibrations for population I stars, and the variation of
colours with metallicity. This was done in two steps: (i) By assuming that the
zero-point calibrations of the various theoretical colors may be in error and
the IRFM yields the right temperatures, we replaced
\teff's of the K92 models with those \teff's that give the same value of the
colours but using the empirical calibrations obtained from population I stars:
this coordinate transformation corresponds to the application of systematic
corrections, function of \teff, to the K92 temperatures, equal to the
differences between the empirical and theoretical calibrations for solar
metallicities. (ii) We then replaced the colours of K92 models for solar
metallicity at each temperature, with the colours of K92 models for the
approximate input value of the metallicity; these colours were obtained by
means of a cubic polynomial interpolation through theoretical colours given by
K92; this second coordinate transformation corresponds to the application of
systematic corrections, still a function of \teff, to the observed colours to
consider the individual metallicity of the star under consideration. Note that
this step implies that though the K92 models may give uncorrect colours at a
given temperature, they correctly predict the dependence of colours on metal
abundances. This last assumption was made also by K93. 
\item We applied the previous transformations separately for both giants 
and dwarfs; when comparing theoretical and empirical calibrations, we assumed 
$\log g=$4.5 for luminosity class V; for luminosity class III, we assumed 
$\log g=$2 for \teff$>4500$~K, and $\log g=($\teff$-3500)/500$ for 
\teff$<4500$~K. These values approximately match the observed $\log g$\ for 
dwarfs and giants respectively; however, this assumption is not critical, since
gravity corrections are generally small. The best temperature for each colour
as function of gravity 
was then derived by linear interpolation/extrapolation between the values
obtained for the two different logarithmic gravities, using the approximate
input value for the stellar gravity. 
\item We then averaged temperatures obtained from each colour, assigning
weight 4 to $V-K$, 0.5 to $V-R$, and 1 to all other colours; these weights were
attributed on the basis of the observed residuals about the fitting polynomials
used for population I stars. 
\item We used this input temperature (and the initial values of gravity, 
metal abundance and microturbulent velocity) to iterate the abundance analysis
until (i) we had no trend of the abundances derived from individual Fe~I 
line with expected line strength (varying \vt); (ii) the value of the model
abundance was identical to the derived Fe abundance (varying [A/H]); and (iii)
the abundances derived from Fe~I and Fe~II lines were the same (varying
$\log g$). This abundance analysis was done using model atmospheres extracted
from the grid of K92. 
\item Finally, we repeated the whole procedure of temperature derivation
entering the new values of gravity, metal abundance, and microturbulent 
velocity obtained after the previous step, and iterated this procedure until 
we converged to a consistent set of atmospheric parameters\footnote{ Note that
there is a small inconsinstency here, since we used model atmospheres computed
with a microturbulent velocity of 2~\kms.}. No iteration was 
usually required when $V-K$\ colours were available, since they are only weakly 
dependent on gravity and metallicity; one or two iterations were required for 
the other cases.
\end{enumerate}

\section{ Atmospheric parameters for reanalyzed stars }

For all program stars, new values for the atmospheric parameters were derived,
following the iterative procedure described in Sect.~3. We used the $EW$s
from the various sources cited in Sect.~2, and photometric data from Hauck
\& Mermilliod (1990), Schuster \& Nissen (1988), Twarog \& Anthony-Twarog
(1994), Laird et al. (1988), Stone (1983), Pilachowski (1978), Arribas \&
Martinez-Roger (1987), and Alonso et al. (1994); however, only $b-y$\ colours
were used for E93's stars, and $b-y$\ and $V-K$\ colours for ZM's stars.
Reddenings for stars considered in GS1 and GS2 were taken from these papers;
those for the SKPL giants were taken from Twarog \& Anthony-Twarog (1994),
while no reddening was assumed for the dwarfs observed by TLLS, E93, and ZM90. 

Since $EW$s for very few lines of Fe~I were usually available from the studies
of TLLS and SKPL, \vt's were obtained using a mean relation drawn from our
program stars (\vt=$-0.322\,\log g+2.22$~\kms), except for a few cases
(generally very cool stars) in which obvious trends with line strength were
present; based on the dispersion along this mean relation, we estimate that
errors of this \vt's are $\pm 0.3$~\kms. For the warmer stars considered by
E93, a dependence on \teff\ must be included. We find that a good representation
is given by the relation (\vt=$1.19\,10^{-3}\,$\teff$-0.90\,\log g-2$~\kms). 
We also revised the Fe~I $gf$'s, to put them in a scale consistent with
that adopted in this paper (see Carretta et al. 1996). 

The derived atmospheric parameters are listed in Tables~2-6 (available in
electronic form). 

\begin{table}
\caption{ Atmospheric parameters for stars in the original sample }
\label{tab:2}
\end{table}

\begin{table}
\caption{ Atmospheric parameters for stars in the TLLS sample }
\label{tab:3}
\end{table}

\begin{table}
\caption{ Atmospheric parameters for stars in the SKPL sample }
\label{tab:4}
\end{table}

\begin{table}
\caption{ Atmospheric parameters for stars in the E93 sample }
\label{tab:5}
\end{table}

\begin{table}
\caption{ Atmospheric parameters for stars in the ZM90 sample }
\label{tab:6}
\end{table}

\section {Discussion of the adopted parameters}

\subsection { Kurucz 1992 and Kurucz 1995 model atmospheres }

After the draft of this paper was ready, we were aware that the convective flux
in Atlas 9 version used to compute the 1992 models stored on the CD-ROM 13
generated discontinuities in the grids of the colour indices for \teff
in the approximate range between 6700~K (for $\log g=2$) to 8000~K (for $\log
g=4.5$). The convection formalism was improved (Castelli, 1996) and
Kurucz (1995) recomputed most of the cool model atmospheres 
(K95 models). The 1995 models differ from the 1992 models mostly for an improved
convection, for the larger number of layers (72 rather than 64) which extend toward
lower optical depths, for a better treatment of the radiation emerging from the
uppermost layers, and for a few changes in some opacity routines, as that for
H$^-$. Furthermore, all the colours are recalibrated on a different ATLAS9 Vega
model (Castelli \& Kurucz 1994).
Rather than redoing all our lengthy computations, we directly compared
colour indices and abundances from the K92 and K95 models, made recently
available to us by R. Kurucz. First we considered colors, and in
particular Johnson $B-V$\ and $V-K$, and Str\"omgren $b-y$\ (these are the
colors having the largest weight in our \teff\ determinations). We found small
constant offsets of 0.004, 0.012 and 0.025 for $b-y$, $B-V$, and $V-K$\
respectively between the original K92 colors and those we obtained
using K95 models (in the sense that K95 colors are redder). These small
offsets are due to slightly different assumptions about the atmospheric
parameters for the reference star Vega. For the stars used in this series of
papers, the theoretical \teff-scale obtained using K95 models is $\sim 30$~K
warmer. For all colors, there is a peak in the color residuals (amounting to
$\sim 0.02$, $0.015$, and $0.045$\ for $b-y$, $B-V$, and $V-K$\ respectively,
again K95 colors being redder) over a small range of \teff\ ($\sim 6750$\ for
$\log~g=2$, and $\sim 8000$~K for $\log g=4.5$). The peak residual is smaller
for metal-poor model atmospheres. Given the small
\teff\ range where these larger corrections (corresponding to 30-120~K,
depending on the colour used) apply, they have negligible impact on the
polynomial fitting curves used in our semiempirical procedure. Furthermore, the
peaks fall outside the \teff range for stars considered in this series (and in
the parallel paper on RR Lyrae at minimum light by Clementini et al. 1995). 

We then compared abundances obtained using K92 and K95 model atmospheres for a
few typical cases, adopting in both cases the same set of atmospheric
parameters. We found that K95 atmospheres yield larger abundances (by 0.008
dex) than K92 atmospheres for low excitation lines of easily ionized elements;
and smaller abundances (by 0.003 dex) for high excitation lines of dominant
species (like the OI IR triplet). 

On the whole, we regard differences between results obtained using K92 and K95
model atmospheres as negligible with respect to other sources of error in our
analysis, and in this series of papers we will keep the results obtained using
K92 atmospheres. 

\subsection {Trends of Fe abundances with excitation potential}

Use of K92 atmospheres and of the new sets of atmospheric parameters allows an 
homogeneous comparison to solar abundances; however, we noticed some 
inconsistencies, that are discussed in this section.

Dalle Ore (1992) found a systematic trends of abundances from individual Fe~I
lines with excitation potential for metal-poor giants, in the sense that
temperature derived from line excitation is much lower than that derived from
colours (that is, a negative slope $\delta\theta$\ of Fe abundances with
excitation potential). We found a similar trend using star in our original
sample, that admittedly included a few stars; however the large spectral
coverage allowed derivation of $EW$s for a large set of lines having accurate
laboratory $gf$'s. We found the following average values for the slope
$\delta\theta$\ in the excitation-abundance plane: 
\begin{itemize}
\item all stars: $-0.047\pm 0.009$~dex/eV ($\sigma=0.041$~dex/eV, 19
stars) 
\item $\log g>4$: $-0.038\pm 0.005$~dex/eV ($\sigma=0.013$~dex/eV, 6
stars) 
\item $3<\log g<4$: $-0.052\pm 0.016$~dex/eV ($\sigma=0.040$~dex/eV,
6 stars) 
\item $\log g<3$: $-0.050\pm 0.022$~dex/eV ($\sigma=0.058$~dex/eV, 7
stars) 
\end{itemize}
On the whole, the value of the slope seems to be weakly dependent on surface 
gravity, while the scatter seems to be a function of gravity. This suggests 
that either the effective temperatures of giants are ill-defined (e.g. due to 
errors in the estimate of reddening), or that the atmospheres of giants are 
somewhat different one from the other, or both. On the other side, the average
slope for stars in the large sample by E93 (all with $\log g>3$) is quite
different ($\delta\theta=+0.027\pm 0.002$, $\sigma=0.025$, 203 independent
estimates for 187 stars); a clear trend with overall metal abundance is present:
$\delta\theta=(0.051\pm 0.005){\rm [Fe/H]}+(0.040\pm 0.021)$. However, the
determinations of the slope using stars in the E93 sample are based on a few
lines with solar oscillator strengths. Contamination by blending features may
cause spurious trends. This point should then be reexamined using an extended
line list for a large sample of stars. 

\begin{figure}[htbp]
\psfig{figure=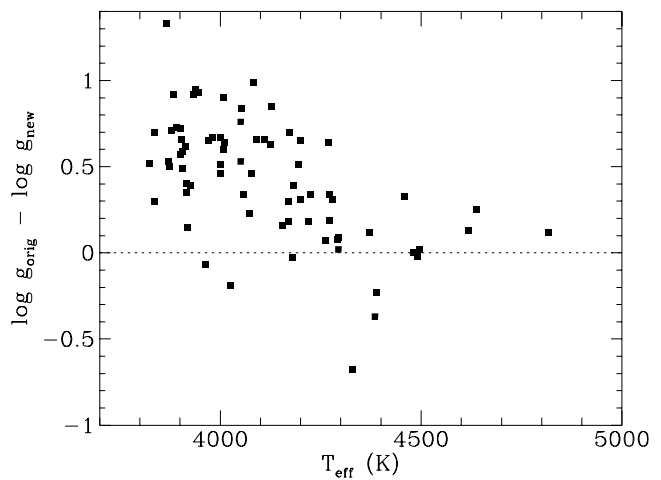,width=8.8cm,clip=}
\caption{Differences between gravities derived from the position in the c-m
diagram ($\log g_{\rm orig}$) and from Fe equilibrium of ionization ($\log
g_{\rm new}$) as a function of \teff\ for the globular cluster giants
considered by Carretta \& Gratton (1996). A mass of 0.8~$\Msol$\ was adopted for
these stars}
\label{fig:2}
\end{figure}

\begin{figure}[htbp]
\psfig{figure=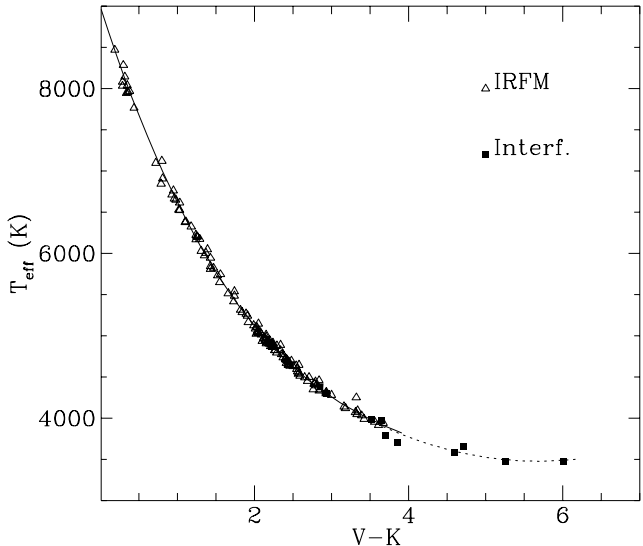,width=8.8cm,clip=}
\caption{Comparison between the ($V-K$)-\teff\ calibrations obtained using
\teff's from IRFM and from diameters measured interferometrically (Di Benedetto
\& Rabbia 1987)}
\label{fig:3}
\end{figure}

\subsection {Gravities and Fe equilibrium of ionization}

For another program (Carretta \& Gratton 1996), we applied this same procedure
to the analysis of the spectra of globular cluster giants, where gravities can
be determined from luminosities, masses and effective temperatures. That
analysis showed that the difference between abundances given by neutral and
singly ionized lines is a function of \teff. This is illustrated by the data in
Fig.~\ref{fig:2}, where we plotted the differences between the gravities for
globular cluster giants obtained from the procedure described in this paper,
and those derived from the position in the c-m diagram, assuming a mass of
0.8~$\Msol$. While the agreement between the two sets of gravities is fairly
good for \teff$>4600$~K, gravities derived from the Fe equilibrium of
ionization are on average too low for stars cooler than 4400~K. 

In the following we will consider three possible explanations for this
discrepancy: 
\begin{enumerate}
\item The IRFM cannot be reliably used for M-stars from ground-based
observations alone, due to the strong molecular bands in the near-IR spectral
region. It is then possible that our empirical calibration for population I
stars, based on the application of the IRFM method to ground-based
observations, is not accurate for the coolest stars. On the other side, \teff's
for M giants can be derived from determinations of stellar angular diameters.
We then tested the cool end of our \teff-colour calibrations by comparing the
\teff's derived using the IRFM with those that can be obtained from
interferometric determinations of stellar diameters (Di Benedetto \& Rabbia,
1987: see Fig.~\ref{fig:3})\footnote{We found that these determinations give a
scatter much smaller than the \teff's from lunar occultations (Ridgway et al.
1980) in colour vs \teff's diagrams}. We found however an excellent agreement
between the two sets of \teff's; a calibration obtained by merging the two sets
differs by $\leq 20$~K from that obtained using \teff's from the IRFM alone for
\teff$>4000$~K. 
\item Some Fe overionization is expected for the coolest, very low
gravity giants. We performed an explorative statistical equilibrium computation
for the coolest star in our own sample (HD~187111) using MULTI code (Scharmer
\& Carlsson, 1985; Carlsson 1986) and a 60-level Fe~I model atom (for a
full description of this model atom and the methods used in this and other
statistical equilibrium computations, see Gratton et al. 1996). We parametrized
the poorly known cross sections for collisions with H~I atoms by matching
observations of RR Lyrae at minimum light (see Clementini et al. 1995). We found
that the maximum non-LTE correction to the Fe~I abundance in HD~187111
compatible with the observations of RR Lyrae variables (where departures from
LTE are expected to be larger than in the program stars) is 0.08~dex
(abundances from Fe~II lines are not influenced by departures from LTE). A
smaller correction (0.04~dex) is expected for HD~136316, while non-LTE
corrections to Fe abundances should be very small for the other program stars.
If LTE abundances from Fe~I lines for HD~187111 are corrected for this
amount, we derive slightly larger surface gravities ($\sim 0.25$~dex).
This correction is about half the value required to reduce the luminosity of
this metal-poor star below that expected at the tip of the red giant branch.
Hence, departures from LTE are a promising candidate to explain part of the
observed discrepancy. 
\item We finally tested the hypothesis that the model atmospheres may be
responsible for the observed discrepancies. This was done by replacing BEGN
models to the K92 ones for a few typical cases. Abundances for the Sun derived
from the K92 models and updated laboratory $gf$'s are close to those given by
the Holweger \& Muller (1974) model, and to the meteoritic values (Anders \&
Grevesse 1989), while it is well known that abundances provided by BEGN solar
model are too low by $\sim 0.08-0.15$~dex (the exact value depends on the line
list used). Furthermore, K92 model reproduces the solar flux distribution much
better than the BEGN model; hence K92 models should be preferred in the
analysis of solar type stars. However, for all the species investigated, the
K92 and BEGN models yield stellar abundances relative to the solar ones which
differ less than 0.03 dex for \teff$>4400$~K. When doing a relative analysis
for cooler stars, the ionization equilibrium given by BEGN model atmospheres is
in better agreement with that obtained using \teff's from IRFM and gravities
from the c-m diagram, although the discrepancy is not completely canceled.
Gravities obtained from the equilibrium of ionization are too low when using
K92 models for these cool stars. This fact suggests that K92 models are not
fully adequate for the analysis of stars with \teff$<4400$~K. 
\end{enumerate}

Since in the present series of papers we are mainly concerned with stars with
\teff$>4400$~K, we will use K92 model atmospheres; however, results for the
coolest stars require further investigations: whenever possible, conclusions
will only be drawn using warmer stars (preferably dwarfs). 

\begin{figure*}
\psfig{figure=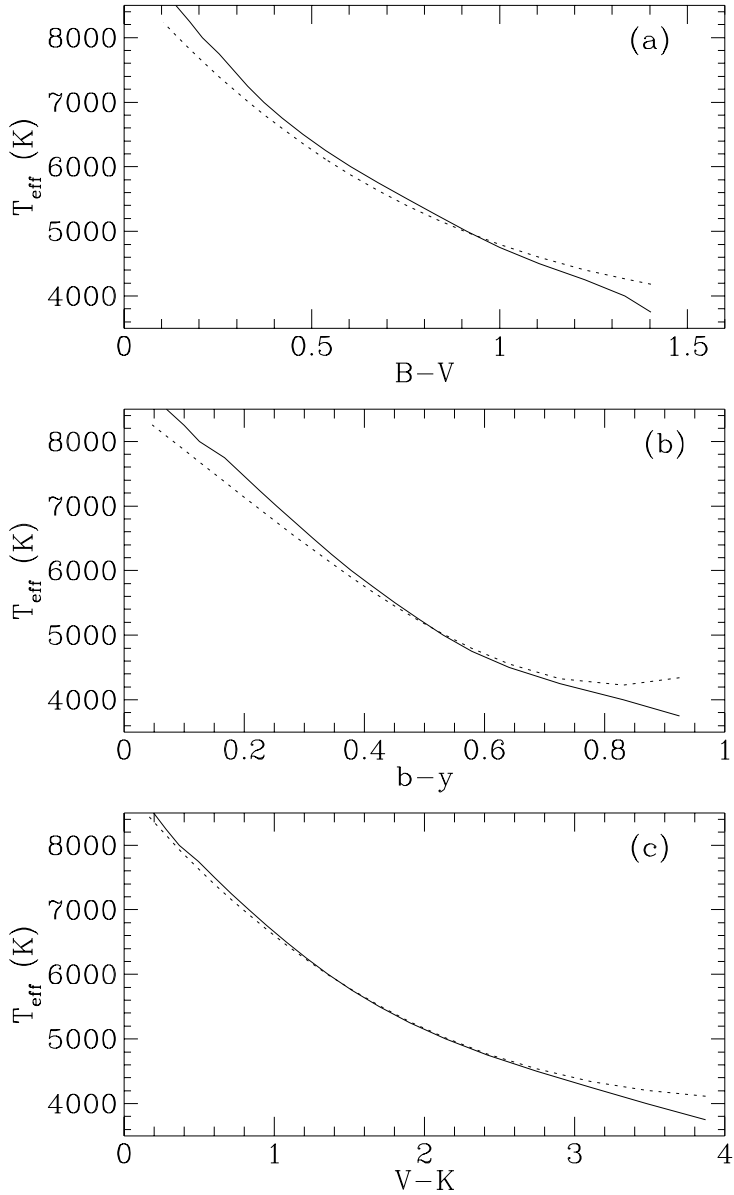,width=10cm,clip=}
\caption[ ]{ Comparison between the present empirical \teff-colour calibrations
for population I dwarfs (dashed line) and the theoretical calibrations directly
based on K92 fluxes (Kurucz 1992) (solid line) for various colours. For brevity,
only the comparisons for $B-V$\ (panel a), $b-y$\ (panel (b), and $V-K$\ (panel
c) are shown in this figure} 
\label{fig:4}
\end{figure*}

\begin{figure*}
\psfig{figure=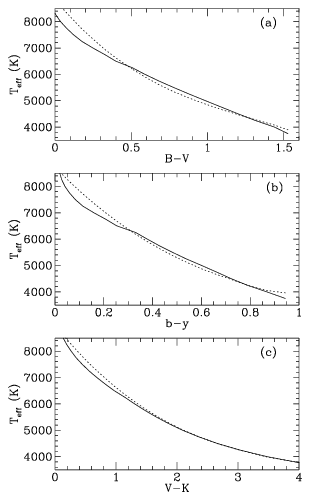,width=10cm,clip=}
\caption[ ]{ The same as Fig.~4 but for population I giants} 
\label{fig:5}
\end{figure*}

\section { Comparison between empirical calibrations and theoretical colours }

Figures~\ref{fig:4} and ~\ref{fig:5} compare the present empirical \teff-colour
calibrations for population I dwarfs and giants respectively with the
theoretical calibrations directly based on K92 fluxes for the Johnson $B-V$,
$V-R$, $R-I$, $J-K$, and $V-K$\ colours, and for the Str\"omgren $b-y$\ index.
The agreement between the empirical and theoretical calibrations is very good
for $V-K$; significant deviations can be noticed only at very low \teff's
(\teff$<4500$~K) for dwarfs, and high \teff's (\teff$>6000$~K) for giants,
where we have only a few calibrating stars. The comparison is also quite good
for $B-V$, $R-I$, and $b-y$, although corrections (a few hundredths of mag) are
not negligible. In particular, we notice that the $B-V$\ colour for the Sun
derived from our empirical calibration ($B-V=0.62$) is 0.03~mag bluer than that
provided by the theoretical calibration. Similar corrections should be applied
when colours from K92 models are applied to theoretical isochrones. The
comparison is poorer for the $V-R$\ colour. 

\section { Comparisons with other \teff-scales }

Of particular interest is the comparison of the current \teff-scale with those 
recently used by other authors for metal-poor stars. 

\subsection { Giants }

\subsubsection { Dalle Ore et al. (1996) }

Dalle Ore et al. (1996) used the K92 models in an analysis of the chemical
composition of HD~122563; their \teff\ (4590~K) derived with the IRFM
agrees very well with our value for this star (4583~K). 

\begin{figure}[htbp]
\psfig{figure=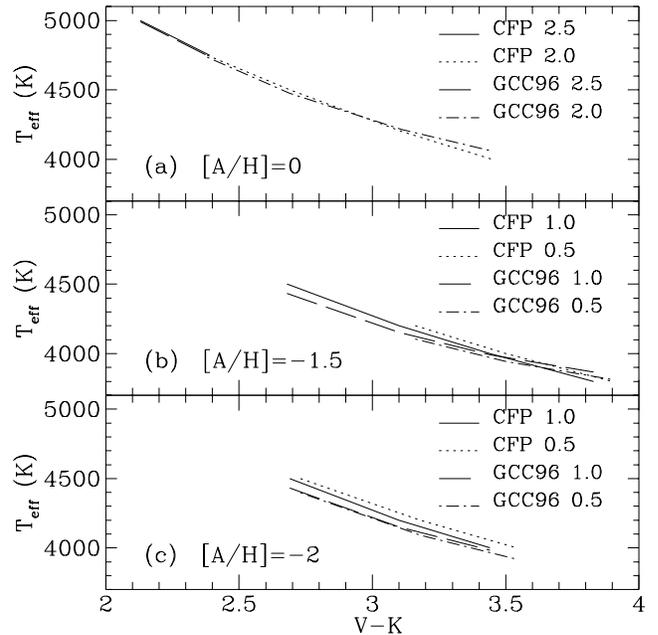,width=8.8cm,clip=}
\caption{Comparison between the present \teff-colour calibrations for giants
(GCC96) and the calibrations by Cohen et al. (1978: CFP). Panel a shows results
for [Fe/H]=0, panel b for [Fe/H]=-1.5, and panel c for [Fe/H]=-2. Different
line types are used for models with different values of $\log g$}
\label{fig:10}
\end{figure}

\subsubsection { Cohen et al. (1978) }

Figure~\ref{fig:10} compares the present calibration of $V-K$\ with that obtained by
Cohen et al. (1978), often used in the analysis of globular cluster giants.
Cohen et al. scale is based on $JHK$\ magnitudes and colours, calibrated
against the old Kurucz (1979) model atmospheres: it predicts that $V-K$\
colours are independent of gravity and metal abundance. The three panels of
Fig.~\ref{fig:10} compare the two scales in different abundance regimes and for
different values of gravities. While the overall comparison is fairly good, the
metallicity dependence is different in the two cases, since we are using the
K92 models. Furthermore, the gravity dependence at low \teff's and
metallicity seems not negligible; however, it is now clear that K92 models are
not fully adequate in the analysis of the coolest, very metal-poor giants. 

\begin{figure}[htbp]
\psfig{figure=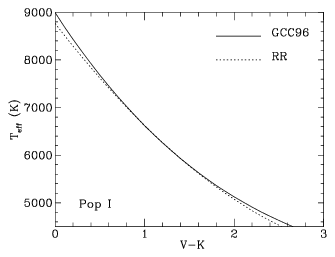,width=8.8cm,clip=}
\caption{Comparison between the \teff\ scale of this paper (GCC96) and that
of Clementini et al. (1995: RR) for population I dwarfs}
\label{fig:11}
\end{figure}

\subsubsection { Clementini et al. (1995) }

It is useful to compare our \teff\ scale with that adopted by Clementini et al.
(1995), since results of the abundance analysis in that paper will be used in
the second paper of this series to calibrate our own statistical equilibrium
computations. Figure~\ref{fig:11} compares the two empirical calibrations of the
$V-K$\ index: the relations are virtually identical in the typical temperature 
range for RR~Lyrae at minimum light (6000-6300~K).

\begin{figure}[htbp]
\psfig{figure=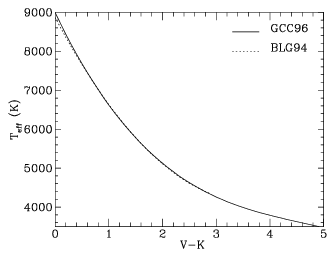,width=8.8cm,clip=}
\caption{Comparison between the \teff scale of this paper (GCC96) and that of
BLG; the agreement is excellent, but the range of validity of the BLG scale is
smaller than the present scale}
\label{fig:12}
\end{figure}

\subsubsection { Blackwell \& Lynas Gray (1994) }

The comparison between the \teff\ scale of this paper and that of BLG is shown
in Fig.~\ref{fig:12}: the agreement is excellent, but the range of validity of
the BLG scale is smaller than the present scale. 

\begin{figure}[htbp]
\psfig{figure=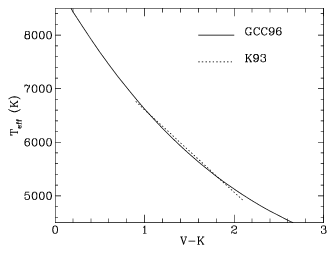,width=8.8cm,clip=}
\caption{Comparison between the \teff\ scale of this paper (GCC96) and K93 
for population~I dwarfs}
\label{fig:13}
\end{figure}

\subsection { Dwarfs }

\subsubsection { King (1993) }

More intriguing is the situation for metal-poor dwarfs. The present \teff\
scale agrees fairly well with that of K93, as shown by the comparisons of
Fig.~\ref{fig:13}; this is not surprising, since they were derived using the
same model atmospheres and a similar procedure. 

\begin{figure}[htbp]
\psfig{figure=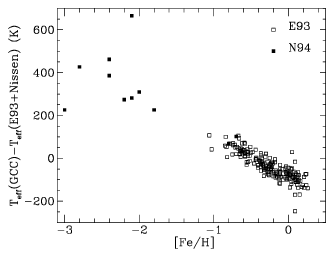,width=8.8cm,clip=}
\caption{Plot of the differences between original \teff's for stars in the 
Edvardsson et al. (1993) and Nissen et al. (1994) samples and those derived 
using our calibration, as a function of our value for [Fe/H]}
\label{fig:14}
\end{figure}

\begin{figure}[htbp]
\psfig{figure=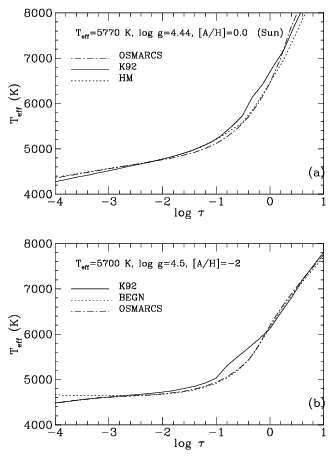,width=8.8cm,clip=}
\caption{$T(\tau)$\ relations for the Sun (panel a) and a metal-poor dwarf 
(panel b) using different model atmospheres: K92 (solid line), OSMARCS 
(dot dashed line), Holweger \& M\"uller (1974: HM, dotted line in panel a),
and BEGN (dotted line in panel b)}
\label{fig:15}
\end{figure}

\subsubsection { Edvardsson et al. (1993) }

Recently, E93 published the results of the analysis of a very extensive
spectroscopic survey of field dwarfs. All stars considered by E93 have
[Fe/H]$>-1$; however, Nissen et al. (1994) presented the results of the
analysis of a few more metal-poor dwarfs whose atmospheric parameters were
derived with a similar technique. A star-to-star comparison between our and E93
and Nissen et al. sets of atmospheric parameters shows large discrepancies in
the adopted \teff's for a few stars. This effect is systematic, as shown by
Fig.~\ref{fig:14}, which displays the difference between the \teff's derived by
the two techniques for the stars observed by E93 and Nissen et al. against
[Fe/H]. The best fit regression line with [Fe/H] is: 
\begin{equation}
\nonumber
\Delta T_{\rm eff}=-(174.9\pm 7.5){\rm [Fe/H]}-(81\pm 30),
\end{equation}
based on 187 stars. The Persson linear regression coefficient is $r^2=0.746$,
but there are minor, not negligible correlations with gravity and \teff; the
regression using all three variables is: 
\begin{eqnarray*}
\Delta T_{\rm eff}&=&(-171.5\pm 7.2){\rm [Fe/H]}-(45\pm 14)\log g \\
~                 &~&- (0.0492\pm 0.0074) T_{\rm eff}+(405\pm 27),
\end{eqnarray*}
with $r^2=0.798$, a marginal but significative improvement. This comparison
shows that \teff's may be different by as much as $\sim 400$~K for the most
metal-poor stars, even though the same colours are used; this is an enormous
difference causing discrepancies as large as $0.4\div 0.5$~dex in the derived
[Fe/H]. A small part of this discrepancy can be attributed to the rather large
values of reddening $E(b-y)$\ adopted by Nissen et al. (1994), and for
individual stars there are colour indices yielding slightly different \teff's
using our prescriptions. However most of this discrepancy is real, and at
{\it prima facie} surprising, since both \teff\ scales are linked to accurately
determined \teff's for Population I stars, for which a good agreement exists
among determinations by different authors. However, a major difference is the
use of different grids of model atmospheres: E93 and Nissen et al. used the new
(unpublished) OSMARCS model atmospheres, while we used the K92 ones. Even
though these two sets of atmospheric models use very similar line opacities,
they differ in the treatment of convection, since K92 models include some
overshooting, which is not present in OSMARCS atmospheres. This causes
different $T(\tau)$\ relations, as shown in Fig.~\ref{fig:15} for both the Sun
and a metal-poor dwarf atmospheres; the K92 models have a bump at $\log \tau
\sim 0$, which is not present in OSMARCS models. In metal-rich atmospheres,
this bump is at rather large optical depths, and it does not affect
significantly the stellar colours. Instead, the bump is at shallower optical
depths in the more transparent atmospheres of metal-poor stars; the consequence
is that the two models predict a very different dependence of colours on metal
abundance: e.g. at $b-y=0.35$, OSMARCS models with [Fe/H]=0 are $\sim 210$~K
warmer than models with [Fe/H]=$-1$, while the analogous difference for the K92
models is only $\sim 40$~K. The discrepancy between the two grids is smaller
for low \teff\ stars, but these are not included in the E93 and Nissen et al.
papers. 

The most sensitive test for model atmospheres is the comparison between
predicted and observed solar limb-darkening relations. Blackwell, Lynas-Gray
and Smith (1995) show that this comparison clearly favors the K92 model
atmospheres with respect to the OSMARCS ones; similar results were also
obtained indipendently by Castelli and Gratton (1996) and Edvardsson (private
communication). This fact suggests that at least in the solar atmosphere
convection transports energy more efficiently around unity optical depth than
the mixing-length theory predicts. This may be a consequence of neglecting the
inhomogeneities observed in the solar atmosphere. The larger freedom given by
the addition of a new parameter (related to overshooting) allows a better
representation of the temperature structure of the solar atmosphere in K92
models. Since it is not entirely clear that this also applies to atmospheres
different from that of the Sun (see e.g. Castelli and Gratton 1996), in the
next two subsections we will further investigate the consistency of the present
scale, by comparing our \teff's with those obtained by directly using the IRFM
for metal-poor dwarfs, and the model predictions with observed profiles of the
Balmer lines. We found a good internal consistency in results obtained with
the K92 models.

\subsubsection { \teff's from IRFM for subdwarfs }

Magain (1987) derived \teff's for eleven
metal-poor dwarfs using the IRFM and the BEGN model atmospheres; it must
however be noticed that his \teff's derived from $J$\ and $K$\ magnitudes are
quite different (those derived from $J$\ being lower by $131\pm 12$~K on
average). His \teff-scale is lower than ours by $\sim 140$~K, with no trend
with metal abundance; this difference is equal to the difference between the
\teff-scale of Magain and that of Saxner \& Hammarb\"ack (1985) at
[Fe/H]$=-0.5$, the lower edge of validity of this last scale (see also K93).
However, to make a meaningful comparison, Magain's \teff's should be corrected
for the systematic differences between \teff's derived using the IRFM with BEGN
and K92 model atmospheres. These have been computed only for population~I stars
(BLG; M\'egessier 1994); if these corrections are applied,
Magain's \teff's are increased by 90~K on average (the correction is rather
large for \teff's derived from $J$\ magnitudes, while it is quite small for
those obtained from $K$). The mean difference with the current \teff-scale is
then reduced to $46\pm 14$~K. Furthermore, \teff's derived from $J$\ and
$K$\ magnitudes would now be in fair agreement with each other, the mean
difference being reduced to $21\pm 12$~K. This test should be repeated with
values for the $R$\ ratio of the IRFM appropriate for metal-poor stars:
however, these early results suggest that the dependence of colours on metal
abundance given by K92 models should not be far from correct.

\begin{figure}[htbp]
\psfig{figure=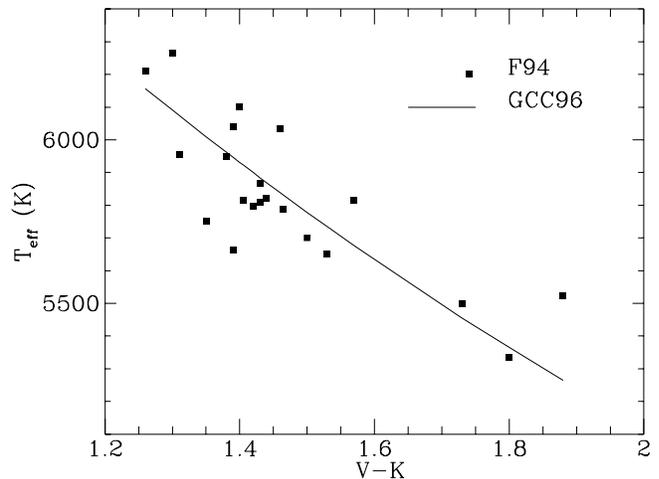,width=8.8cm,clip=}
\caption{Comparison of our \teff\ calibration (GCC96) with \teff's deduced
from the wing of Balmer lines (Fuhrmann et al. 1993, 1994: F94) for some
metal-poor dwarfs}
\label{fig:16}
\end{figure}

\subsubsection { \teff's from Balmer line profiles for subdwarfs }

\begin{figure}
\psfig{figure=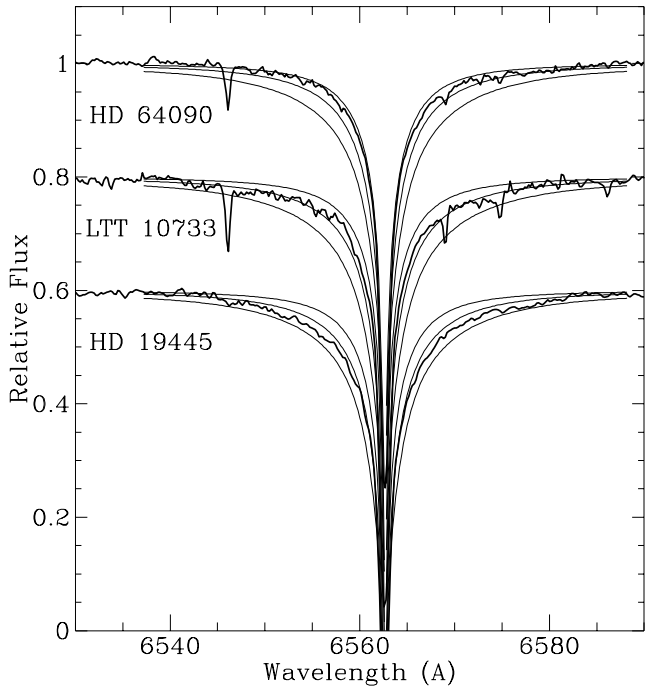,width=8.8cm,clip=}
\caption[ ]{ Comparison between observed H$_\alpha$\ profiles for three
subdwarfs (HD~64090, LTT~10733, and HD~19445) and synthetic profiles computed
with \teff=5500, 6000, and 6500~K. In these computations, the metal abundances
were set at [A/H]=-1 for LTT10733, and [A/H]=-2 for the two other stars }
\label{fig:17}
\end{figure}

\begin{table}
\caption{\teff's from H$_\alpha$\ for seven subdwarfs}
\label{tab:8}
\vskip 12pt
\begin{tabular}{lcccccc}
\hline\hline
Star     &$(V-K)$&$[$Fe/H$]_{\rm SN}$&\teff($V-K$)&\teff(H$_\alpha$)\\
\hline
\\
HD~19445 & 1.37~  & -1.91 & 6112 & 6250\\
HD~64090 & 1.72~  & -1.69 & 5510 & 5500\\
HD114762 & 1.41~  & -0.87 & 5964 & 5750\\
HD194598 & 1.35~  & -1.11 & 6077 & 6100\\
HD201891 & 1.39~  & -1.08 & 6009 & 6000\\
LTT10733 & 1.525  & -1.04 & 5788 & 5900\\
LTT11819 & 1.79~  & -1.56 & 5406 & 5400\\
\\
\hline \hline\\
\end{tabular}
\end{table}

While the far wings of H$_\alpha$\ are quite independent from gravity,
metallicity and convection of the
model atmospheres (and are then good temperature indicators for solar type
stars), those of the other Balmer lines are strongly dependent on how
convection is handled when computing model atmospheres, due to its effect on
the temperature stratification of the model. Fuhrmann et al. (1993) employed
this dependence to determine the best value for the pressure scale height in
the mixing length formalism, to be used when modelling atmospheres for
metal-poor dwarfs. Unfortunately, they used the old Kurucz (1979) model
atmospheres, and their results cannot be directly used in the present context.
We note, however, that our \teff-scale compares quite well with the \teff's
determined by Fuhrmann et al. (1994; see Fig.~\ref{fig:16}). 

For another program, we acquired spectra of several subdwarfs at a resolving
power of $\sim 15,000$\ and $S/N\sim 250$\ using the REOSC Echelle 
spectrograph at
the 182~cm Copernicus reflector of Asiago Observatory. This spectrograph uses a
79~gr/mm echelle grating and a large format front illuminated Thomson CCD
detector, with no appreciable diffraction fringes. These spectra are then quite
well suited for the determination of \teff's from the H$_\alpha$ profiles.
Preliminary reduced spectra for seven subdwarfs were used by Clementini et al.
(1995) to derive \teff's from H$_{\alpha}$\ profiles using K92 model
atmospheres. Details about data reduction are given in Clementini et al.: here
we only remind that particular care was devoted to flat fielding and
subtraction of the telluric lines. For two stars (HD~19445 and HD~114762) we
could directly compare the H$_{\alpha}$\ profiles derived from our spectra with
those by Fuhrmann et al. The agreement is excellent: profiles from the two 
sources agree within $\sim 0.5$\%. 

The observed H$_{\alpha}$\ profiles were compared with synthetic profiles
computed using K92 models and our own spectral synthesis code, which included
Doppler, natural damping, resonance (Ali \& Griem 1965, 1966), and Stark (Vidal
et al. 1970) broadening, following prescriptions similar to those adopted by
Fuhrmann et al. (1993, 1994). These prescriptions should be correct for
electron densities above 10$^{11}$ electrons per cm$^3$, that is throughout
most of dwarf atmospheres. However, line formation in the outer part of the
atmospheres may be affected by appreciable deviations from LTE: hence these
part of the profiles (as well as those contaminated by other lines) were not
considered in the \teff\ derivations. A comparison with the solar flux spectrum
(Kurucz et al. 1984) showed that our computed profiles with the K92 model
atmospheres reproduce observations very well (incidentally, good agreement is
also obtained when using the OSMARCS and Holweger \& M\"uller, 1974,
atmospheres). 

A comparison of observed and computed profiles for three stars is given in
Fig.~\ref{fig:17}. Our \teff's from H$_{\alpha}$\ profiles for the seven
subdwarfs are listed in Table~\ref{tab:8}, where metal abundances are from
Schuster \& Nissen (1989)\footnote{ \teff's presently derived from H$_\alpha$\
wings are slightly different from both those obtained in Clementini et al
(1995) and Castelli and Gratton (1996). These small differences with
Clementini et al. are due to the improvement of
our code, the differences with Castelli and Gratton are due to the use made
by them of the BALMER9 and SYNTHE codes from Kurucz. The difference between
the code used in this paper and the Kurucz codes are mostly due to the way
how the transfer equation is solved.}. In this table we also
give \teff's from our photometric calibrations (averaging $V-K$\ colors from
Alonso et al. (1994) and Laird et al. (1988), both transformed into the Johnson
system using the relations given by Alonso et al. 1994). The agreement between
\teff's derived from spectroscopy and photometry is excellent: the mean
difference is $6\pm 43$. The r.m.s. scatter of star-to-star residuals (114~K)
indicate that internal errors are $\sim 100$~K; we attribute these errors to
small ($\sim 1$\%) uncertainties in the location of the continuum level. We
conclude that the comparison with $H_{\alpha}$\ profiles supports our
colour-temperature calibrations. 

\section { Comparison with parameters used in the original analyses }

\subsection { GS1 and GS2 }

The new temperatures and gravities are larger for dwarfs and smaller for giants
with respect to the values used in GS1 and GS2. Metal abundances and
microturbulent velocities are also larger for dwarfs, while they are close to
the old values for giants. We computed linear regression fits through the
residuals between new and old values of the atmospheric parameters as a
function of surface gravity; the relations are as follows: 
\begin{eqnarray*}
T_{\rm eff~us}-T_{\rm eff~old}&=&(45.7\pm 6.1)~\log g_{\rm us}\\
&&-(100\pm 41)~~{\rm K}
\end{eqnarray*}
\begin{eqnarray*}
\log g_{\rm us}-\log g_{\rm old}&=&(0.147\pm 0.027)~\log g_{\rm us}\\
&&-(0.35\pm 0.18)    \\
\end{eqnarray*}
\begin{eqnarray*}
v_{\rm t~us}-v_{\rm t~old}&=&0.088\pm 0.049~\log g_{\rm us}\\
&&-(0.23\pm 0.32)~~{\rm km/s}
\end{eqnarray*}
\begin{eqnarray*}
{\rm [Fe/H]}_{\rm us}-{\rm [Fe/H]}_{\rm old}&=&(0.046\pm 0.011)~\log
g_{\rm us}\\
&&+(0.101\pm 0.076)
\end{eqnarray*}

We notice that the main conclusions of the present series of papers are not
influenced by these variations of the atmospheric parameters, although
abundances for individual stars may be different by as much as 0.3~dex. 

\subsection { TLLS }

These atmospheric parameters are rather different from the original ones; on
average, our \teff's are larger than those found in TLLS by $165\pm 16$~K,
the $\log g$'s by $0.31\pm 0.03$, the [Fe/H]'s by $0.29\pm 0.02$, while the
\vt's are smaller by $0.67\pm 0.03$~\kms. We found no significant trend for
these offsets with \teff, $\log g$, and [Fe/H], except of course for \vt. 

\subsection { SKPL }

Differences with SKPL are much smaller (on average differences in the sense
us-SKPL are $-78\pm 18$~K, $-0.24\pm 0.07$~dex, $0.02\pm 0.02$~dex,
and $-0.12\pm 0.03$~\kms\ for \teff, $\log g$, [Fe/H] and \vt\
respectively), but there are trends with \teff. 

\subsection { E93 }

Significative trends with metal abundances are present when we compare our
adopted parameters for the stars considered by E93 with those adopted in their
preliminary analysis: 
\begin{eqnarray*}
T_{\rm eff,us}-T_{\rm eff,E93}&=&-(180\pm 9){\rm [Fe/H]}_{us}\\
&&-(81\pm 35)~{\rm K},
\end{eqnarray*}
\begin{eqnarray*}
\log g_{us}-\log g_{E93}&=&-(0.31\pm 0.04){\rm [Fe/H]}_{us}\\
&&-(0.11\pm 0.17)~{\rm dex},\\
\end{eqnarray*}
\begin{eqnarray*}
{\rm [Fe/H]}_{us}-{\rm [Fe/H]}_{E93}&=&-(0.165\pm 0.008){\rm [Fe/H]}_{us}\\
&&-(0.024\pm 0.045)~{\rm dex}.
\end{eqnarray*}
These trends may be explained by the differences in the \teff\ scales (see 
Sect.~7.2.2).

\subsection { ZM90 }

No trend with metallicity is present when we compare our parameters with those
originally adopted by ZM90. Mean differences (20 stars) are:
\begin{eqnarray*}
T_{\rm eff,us}-T_{\rm eff,ZM90}=137\pm 4~~{\rm K},
\end{eqnarray*}
\begin{eqnarray*}
\log g_{us}-\log g_{ZM90}=0.65\pm 0.02~~{\rm dex},\\
\end{eqnarray*}
\begin{eqnarray*}
{\rm [Fe/H]}_{us}-{\rm [Fe/H]}_{ZM90}=0.30\pm 0.01~~~{\rm dex},
\end{eqnarray*}
with standard deviations of 20~K, 0.09~dex, and 0.03~dex respectively. The
temperature difference corresponds to the use of the calibration by Magain
(1987) by ZM90. 

\section { Conclusions }

We have presented a new, self-consistent set of atmospheric parameters
for about 300 metal-poor stars, that will be analyzed for the abundances of Fe
and light elements (C, N, O, Na, and Mg) in forthcoming papers of this series.
The most important aspect of this derivation is the determination of a new
\teff\ scale, based on the K92 model atmospheres. Our \teff\ scale is based
on \teff's determined empirically using the IRFM for about 140 population I
stars. We considered separately dwarfs and giants, so that \teff's for stars of
any gravity can be derived by interpolation/extrapolations. Cubic polynomials
drawn through the observational points allows to correct theoretical \teff's
from the K92 models: \teff's appropriate for any star can then be obtained by
an iterative procedure by using the theoretical dependence of colours on metal 
abundance, once abundance determined from the line analysis and gravities
given by Fe equilibrium of ionization are known.

A discussion of our \teff\ scale shows that it gives consistent results for
stars with \teff$>4500$~K, while gravities determined from the equilibrium
of ionization for Fe are too low for giants cooler than this limit. We discuss
several possible causes of this discrepancy: we found that departures from
LTE may explain part of it. However, K92 models are likely not adequate matches
to the atmospheres of the coolest metal-poor giants.

We compared our \teff\ scale with others from the literature. We found
excellent agreement with other \teff\ scales based on the K92 model atmospheres
(in particular, the agreement is good with K93 \teff's), while there is a
serious discrepancy with the \teff's determined for metal-poor dwarfs using the
new OSMARCS models. We attribute this discrepancy to the different way
convection is handled in the two set of models. K92 model atmospheres better
reproduces the solar limb darkening than the OSMARCS models. We find that
additional confirmations of the present \teff\ scale for metal-poor dwarfs are
provided by independent determinations of \teff's obtained by both the IRFM and
the wings of H$_\alpha$. 

\acknowledgements{ We are much thankful to Dr M. Carlsson, who provided a copy
of MULTI code and kindly advised us on its use; to Dr R. Kurucz, who
provided tapes containing his huge line list, and model atmospheres in advance
of publication; and to the referee (dr B. Edvardsson) for his careful reading
and stimulating suggestions. This research has made use of the Simbad database,
operated at CDS, Strasbourg, France. } 

%
%------------- Appendix ---------------
%
%\app{ } 
%
%
%
%
% ---------- References -------------
%

\end{document}